\documentclass[fp,dvipdfmx,twocolumn]{jpsj3}

\usepackage[dvipdfmx]{graphicx}

\usepackage[usenames]{color}

\usepackage{here}

\def\Sec{Sect. }

\def\msp{m_{\rm sp}}

\def\v#1{\mib #1}

\def\stotz{{S_{\rm tot}^z}}

\def\Mspz{{M_{\rm sp}^z}}

\newcommand{\ket}[1]{\left\vert {#1} \right\rangle}

\newcommand{\aver}[1]{\left\langle {#1} \right\rangle}

\newcommand{\kket}[1]{\left\vert\left\vert {#1} \right.\right\rangle}

\title
{
Ground-State Phase Diagram of (1/2,1/2,1) Mixed Diamond Chains with Single-Site Anisotropy
}

\author
{
Kazuo Hida\thanks{E-mail address: hida@mail.saitama-u.ac.jp}
}

\inst
{Professor Emeritus, \\

Saitama University, Saitama, Saitama, 338-8570\\

}

\recdate
{}

\abst
{
The ground-state phases of mixed diamond chains with ($S, \tau^{(1)}, \tau^{(2)})=(1/2,1/2,1)$, where $S$ is the magnitude of vertex spins, and $\tau^{(1)}$ and $\tau^{(2)}$ are those of apical spins, are investigated with the single-site anisotropy $D$ on the $\tau^{(2)}$-site.
The two apical spins in each unit cell are coupled by an exchange coupling $\lambda$. The vertex spins are coupled with the top and bottom apical spins by exchange couplings $1+\delta$ and $1-\delta$, respectively. 
The ground-state phase diagram is determined using the numerical exact diagonalization and DMRG method in addition to the analytical approximations in various limiting cases. The phase diagram consists of a N\'eel ordered phase, a nonmagnetic Tomonaga-Luttinger liquid phase, and quantized and partial ferrimagnetic phases. A region with anisotropy inversion is found where the Ising-like N\'eel phase is realized for the easy-plane anisotropy $D >0$ and the XY-like Tomonaga-Luttinger liquid phase is realized for the easy-axis anisotropy $D <0$ on the $S=1$ sites.
}

\begin{document}

\maketitle

\section{Introduction}

Various exotic quantum phases emerge in low-dimensional frustrated quantum magnets as a result of the interplay of quantum fluctuation and frustration.\cite{intfrust,diep} The conventional diamond chain\cite{Takano-K-S,ht2017} is one of the simplest examples of such systems that can be analyzed thanks to the infinite number of local conservation laws analytically. Extensive experimental studies have also been conducted on the magnetic properties of the natural mineral azurite that is regarded as an example of distorted spin-1/2 diamond chains.\cite{kiku2,kiku3}

On the other hand, the cases with unequal apical spins are less studied. As a simple example of such cases, we investigated the mixed diamond chain with apical spins of magnitude 1 and 1/2, and vertex spins of magnitude 1/2 in Refs. \citen{hida2021} and \citen{kh2024}. We found exotic quantum phases such as an infinite series of ferrimagnetic phases, quantized ferrimagnetic (QF) phases, and partial ferrimagnetic (PF) phases in addition to the nonmagnetic Tomonaga-Luttinger liquid (TLL) phase..

In the present work, we consider the effect of the single-site anisotropy $D$ on the spin-1 sites of this model.
This paper is organized as follows.
{In \Sec 2}, the model Hamiltonian is presented. 
{In \Sec 3}, various limiting cases are examined analytically, and the classical limit is discussed. In \Sec 4, the ground-state phase diagram determined by the numerical calculation is presented for $D=1$ and $D=-1$ as examples of the easy plane and easy axis anisotropies. It is remarkable that there exists a region where the Ising-like N\'eel phase is realized for easy plane anisotropy $D >0$ and the XY-like TLL phase is realized for easy axis anisotropy $D <0$ on the $S=1$ sites. This implies that the effect of $D$ is inverted in this regime. Other properties of each phase are also discussed. 
The last section is devoted to a summary and discussion.

\section{Model}

\begin{figure}

\centerline{\includegraphics[width=7cm]{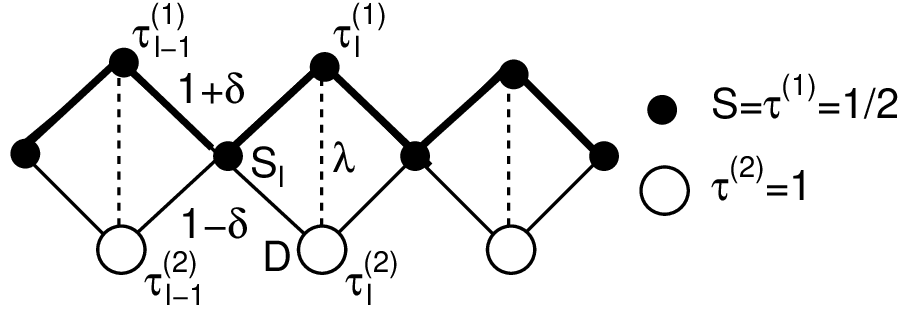}}

\caption{Structure of the diamond chain investigated in this work.}

\label{lattice}

\end{figure}

We investigate the ground-state phases of $(1/2,1/2,1)$ mixed diamond chains with the single-site anisotropy described by the following Hamiltonian:

\begin{align}
{\mathcal H} &= \sum_{l=1}^{L} \Big[(1+\delta)\v{S}_{l}(\v{\tau}^{(1)}_{l}+\v{\tau}^{(1)}_{l-1})\nonumber\\
&+(1-\delta)\v{S}_{l}(\v{\tau}^{(2)}_{l}+\v{\tau}^{(2)}_{l-1})\nonumber\\
&+\lambda\v{\tau}^{(1)}_{l}\v{\tau}^{(2)}_{l}+D{\tau}^{(2)z 2}_{l}\Big],
\label{eq:ham0}
\end{align}
where $\v{S}_{l}, \v{\tau}^{(1)}_{l}$ and $\v{\tau}^{(2)}_{l}$ are spin operators with magnitudes ${S}_{l}={\tau}^{(1)}_{l}=1/2$ and ${\tau}^{(2)}_{l}=1$, respectively. The number of unit cells is denoted by $L$. The total spin $\v{S}_{\rm tot}$, the spontaneous magnetization ${M}^z_{\rm sp}$ and that per unit cell ${m}^z_{\rm sp}$ are defined by
\begin{align}
\v{S}_{\rm tot}&=\sum_{l=1}^L\left[\v{S}_{l}+\v{\tau}^{(1)}_{l}+\v{\tau}^{(2)}_{l}\right],\\
{M}^z_{\rm sp}&=\aver{{S}^z_{\rm tot}},\\
{m}^z_{\rm sp}&=\frac{1}{L}M^z_{\rm sp},
\end{align}
where $\aver{}$ denotes the expectation value in the ground state. 
The lattice structure is depicted in Fig. \ref{lattice}. We consider the region $\lambda \geq 0$ and $1 \geq \delta\geq -1$. The isotropic case ($D=0$) has been investigated in Refs. \citen{hida2021} and \citen{kh2024}.

\section{Limiting Cases}

\subsection{$\lambda=0$}
\label{sec:lambda0}

The system is unfrustrated. In the isotropic case $D=0$, the ground state is the QF phase with spontaneous magnetization $\msp=1$ per unit cell according to the Lieb-Mattis (LM) theorem.\cite{kh2024,Lieb-Mattis} In the presence of easy-axis anisotropy $D < 0$, this state remains as the QF phase with $\msp^z=1$. This phase is called the LM1 phase. On the other hand, in the presence of easy-plane anisotropy $D >0$, the XY component of the magnetization cannot order due to U(1) symmetry. Hence, the ground state turns into the nonmagnetic TLL phase. Hereafter, this phase is called the TLL1 phase.

\subsection{$\delta=1$}
\label{sec:deltap1}

If $\lambda > 0$, the system is unfrustrated. The ground state in the isotropic case is the QF phase with $\msp=1$ according to the LM theorem.\cite{kh2024,Lieb-Mattis} For $D < 0$, this state remains as the QF phase with $\msp^z=1$. We call this phase the LM2 phase to distinguish it from the LM1 phase with a different spin configuration. On the other hand, for $D >0$, the XY component of the magnetization cannot order due to U(1) symmetry. Hence, the ground state turns into the TLL phase. Hereafter, this phase is called the TLL2 phase.

\subsection{$\delta=-1$}
\label{sec:deltam1}

The system is unfrustrated, and the ground state in the isotropic case is the nonmagnetic phase with $\msp=0$ according to the LM theorem.\cite{kh2024,Lieb-Mattis} This phase is expected to be the TLL phase as discussed in Ref. \citen{kh2024}. For $D < 0$, this state turns into the N\'eel phase. On the other hand, for $D >0$, the ground state remains in the nonmagnetic TLL phase. Hereafter, this phase is called the TLL0 phase.

\subsection{$\lambda \gg 1$}
\label{sec:lambdainf}

Hamiltonian Eq. (\ref{eq:ham0}) is decomposed in the following way.

\begin{align}
{\cal H}&={\cal H}_0 + {\cal H}_1,\\
{\cal H}_0&=\sum_{l=1}^L h_l,\\
{\cal H}_1&=\sum_{l=1}^L \left[(1+\delta)\v{\tau}^{(1)}_l(\v{S}_{l}+\v{S}_{l+1})\right.\nonumber\\
&\left.+(1-\delta)\v{\tau}^{(2)}_l (\v{S}_{l}+\v{S}_{l+1})\right],
\end{align}
where
\begin{align}
\label{model3}
{h}_l&=\lambda\v{\tau}^{(1)}_l\v{\tau}^{(2)}_l + D{\tau}^{(2)z2}_l.
\end{align}

The ground states of $h_l$ are given by
\begin{align}
\kket{\frac{1}{2}}_l&=\alpha\ket{\frac{1}{2}}_{1l}\ket{0}_{2l}+\beta\ket{-\frac{1}{2}}_{1l}\ket{ 1}_{2l},\\
\kket{-\frac{1}{2}}_l&=-\left(\alpha\ket{-\frac{1}{2}}_{1l}\ket{0}_{2l}+\beta\ket{\frac{1}{2}}_{1l}\ket{ -1}_{2l}\right),
\end{align}
where
\begin{align}
\alpha&=\sqrt{\frac{1}{2}+\frac{x}{2\sqrt{x^2+2}}},\\
\beta&=-\sqrt{\frac{1}{2}-\frac{x}{2\sqrt{x^2+2}}},
\end{align}
with $x=-1/2 + D/\lambda$. Here $\kket{{T}^z_l}_l$ is the two-spin ground states of $h_l$ with $\tau^{(1)z}_{l}+\tau^{(2)z}_{l}=T^z_l$. The single-spin eigenstate of $\tau^{(\alpha)z}_{l}$ is denoted by $\ket{\tau^{(\alpha)z}_{l}}_{\alpha l}$.

Within the subspace spanned by $\kket{\pm 1/2}_l$, the first-order effective Hamiltonian for the whole chain is given by the spin-1/2 XXZ chain
\begin{align}
{\cal H}^{\rm eff}&=\sum_{l=1}^L \left[J_{\rm eff}^z({T}^z_l(S_l^z+S^z_{l+1})\right.\nonumber\\\
&\left.+\frac{1}{2}J_{\rm eff}^{xy}({T}^+_l(S_l^-+S^-_{l+1}+{\rm h.c.})\right],\label{eq:heff}
\end{align}
with
\begin{align}
J_{\rm eff}^z&\simeq 1-\frac{5\delta}{3}\left(1-\frac{32D}{45\lambda}\right),\\
J_{\rm eff}^{xy}&\simeq 1-\frac{5\delta}{3}\left(1-\frac{4D}{45\lambda}\right),
\end{align}
within the lowest order in $\lambda^{-1}$.

The operators ${T}^{\pm}_l$ are defined so that the relations
\begin{align}
{T}^{\pm}_l\kket{\mp\frac{1}{2}}_l&=\kket{\pm \frac{1}{2}}_l,\\
{T}^{\pm}_l\kket{\pm\frac{1}{2}}_l&=0,
\end{align}
are satisfied.
\begin{table}
\begin{tabular}{ccc}
\hline
parameters& ${\cal H}^{\rm eff}$ & GS\\
\hline
$\delta > \delta_c, D > 0$ & XY-like & TLL2\\
$\delta_c > \delta > 0, D > 0$ & Ising-like AF & N\'eel \\
$\delta < 0, D > 0$ & XY-like& TLL0 \\
$\delta > \delta_c, D < 0$ & Ising-like F & LM2 \\
$\delta_c > \delta > 0, D < 0$ & XY-like & TLL \\
$\delta < 0, D < 0$ & Ising-like AF & N\'eel \\
\hline
\end{tabular}

\caption{Ground-state (GS) phases in the large $\lambda$ limit. $\delta_c=0.6(1+2D/5\lambda)$. F and AF stand for ferromagnet and antiferromagnet, respectively.}
\label{table1}
\end{table}

It is well-known that the ground state of this model is the TLL state for $-1 < \Delta(\equiv J^z_{\rm eff}/ |J^{xy}_{\rm eff}|) \leq 1$, the N\'eel state for $\Delta > 1$, and the ferromagnetic state for $\Delta \leq -1$.\cite{yy3,quantmag} Taking into account that the ferromagnetic phase in (\ref{eq:heff}) corresponds to the ferrimagnetic phase with $\msp^z=1$ in the original model (\ref{eq:ham0}), the ground-state phases of  (\ref{eq:ham0}) for large $\lambda$ are summarized in Table \ref{table1}. The identification of TLL0 and TLL2 phases is made taking into account the continuation to the limiting cases described in \Sec \ref{sec:deltam1} and \Sec \ref{sec:deltap1}, respectively.

It is remarkable that the Ising-like AF phase is realized for easy-plane anisotropy $D >0$ on the $S=1$ sites and the XY-like TLL phase is realized for easy-axis anisotropy $D <0$ on the $S=1$ sites in the range $\delta_c > \delta > 0$. This implies that the effect of $D$ is inverted in this regime. This is also confirmed numerically in the next section. Similar phenomena are reported in various anisotropic frustrated quantum spin-1/2 systems.\cite{okamoto2014,oi2002,tokuno2005,okamoto2002}

\subsection{Classical limit}

The classical phase diagram is not affected by the single-site anisotropy $D$, since it only imposes the constraints on the direction of $\v{\tau}^{(2)}_{l}$ and the classical energy is determined only by the relative angles among other spins and $\v{\tau}^{(2)}_l$. Hence, the classical phase diagram is the same as Fig. 4 of Ref. \citen{kh2024}. It should be noted, however, that the spontaneous magnetizations in LM1 and LM2 phases for $D >0$ are constrained within the XY plane. Hence, they are expected to reduce to the TLL1 and TLL2 phases in the presence of quantum fluctuation.

\section{Numerical Results}

\subsection{Case of easy plane anisotropy : $D=1$}

\begin{figure}[h]

\centerline{\includegraphics[width=6.0cm]{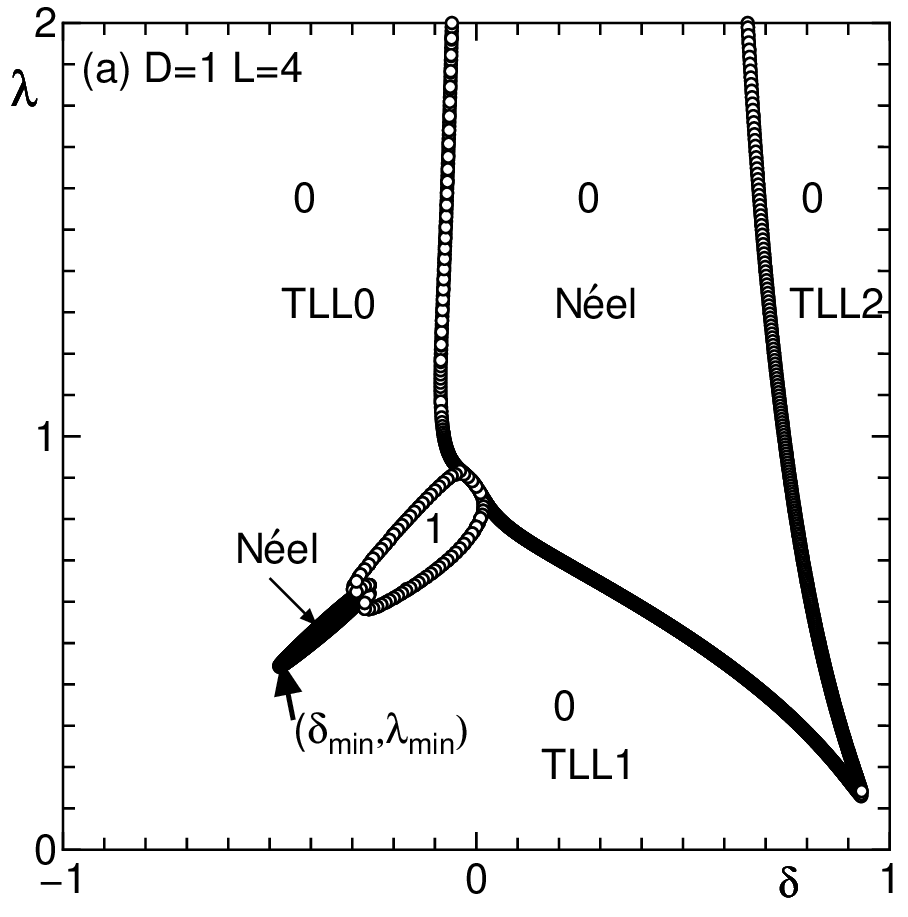}}

\centerline{\includegraphics[width=6.0cm]{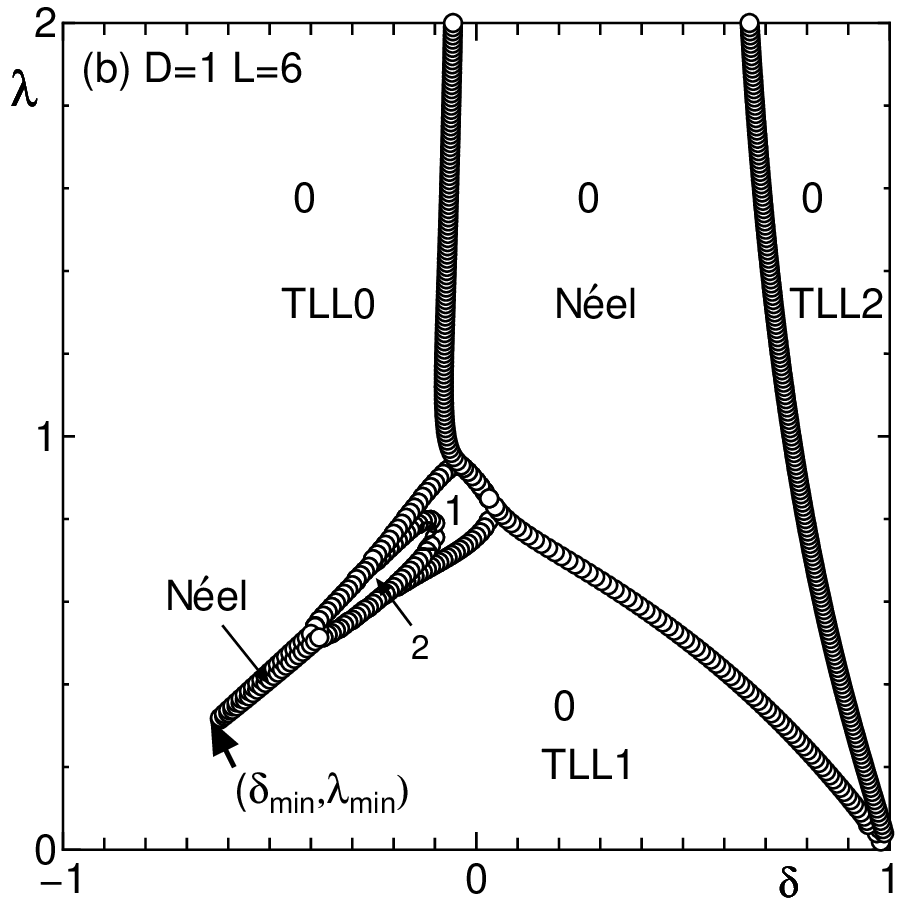}}

\centerline{\includegraphics[width=6.0cm]{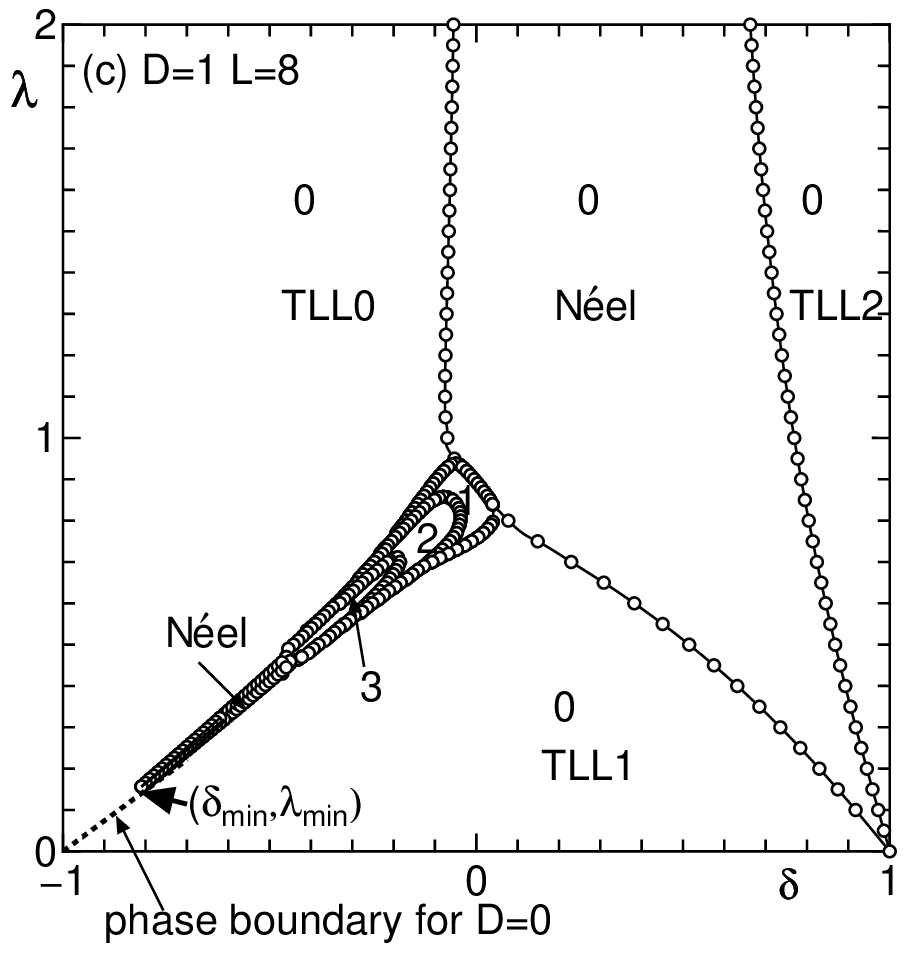}}

\caption{Ground-state phase diagrams based on the NED data with $D=1$ for (a)$L=4$, (b)$L=6$, and (c) $L=8$. The number indicated for each phase is $\Mspz$ of the ground state. The dotted line in (c) is the TLL-LM1 phase boundary for $D=0$ obtained in ref. \citen{kh2024}}

\label{fig:phasep}

\end{figure}

\begin{figure}[h]

\centerline{\includegraphics[width=6.0cm]{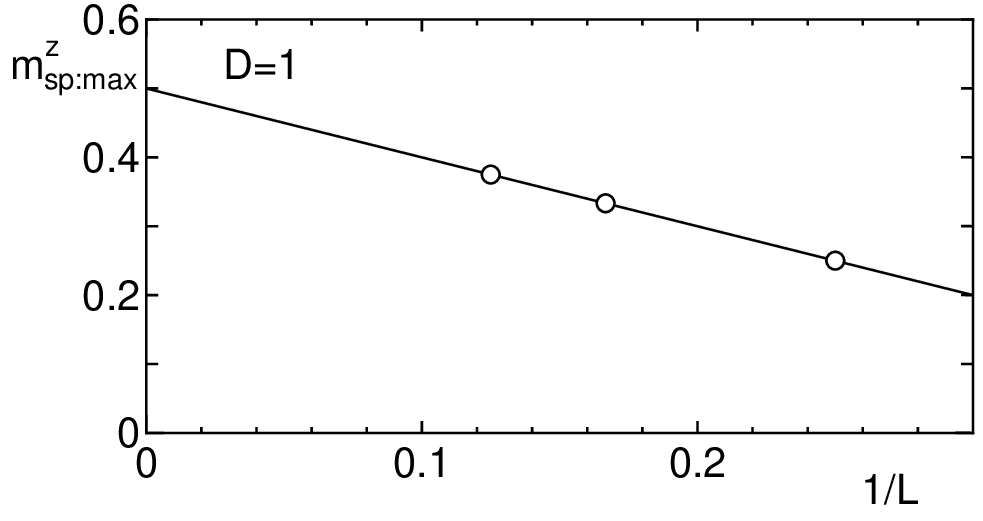}}

\caption{Maximum value of spontaneous magnetization $m^z_{\rm sp:max}$ within the PF phase plotted against $1/L$ for $D=1$. }

\label{fig:spmagmax}

\end{figure}

We have carried out the numerical exact diagonalization (NED) calculation for $L=4$, 6, and 8 to obtain the ground-state phase diagram of Fig. \ref{fig:phasep} for $D=1$. Around the center of the phase diagram, there are several ferrimagnetic phases. In the thermodynamic limit, these phases are expected to form a PF phase in which the spontaneous magnetization varies continuously with parameters $\lambda$ and $\delta$ as in the isotropic case\cite{kh2024}. The maximum value of the spontaneous magnetization per unit cell $m^z_{\rm sp:max}$ within the PF phase is plotted against $1/L$ in Fig. \ref{fig:spmagmax}. This result suggests that $m^z_{\rm sp:max}$ tends to the value close to 0.5 in the thermodynamic limit.

The N\'eel and TLL phases are distinguished by the value of $\stotz$ of the lowest lying excitation. In the N\'eel phase, the lowest excitation has $\stotz=0$, while it has $\stotz=1$ in the TLL phase. 
The spontaneously dimerized phase also has the lowest excitation with $\stotz=0$. However, the presence of this phase can be excluded by considering the spin inversion parities of the ground state and lowest excited state. In the N\'eel phase, the ground and first excited states of a finite chain have opposite parity, since one of them is a symmetric superposition of two N\'eel ordered states that are related by spin inversion, and the other is their antisymmetric superposition. On the other hand, the spin inversion does not change the spontaneously dimerized state. The latter behavior is not found within the present numerical calculation.

\begin{figure}[h]

\centerline{\includegraphics[width=6.0cm]{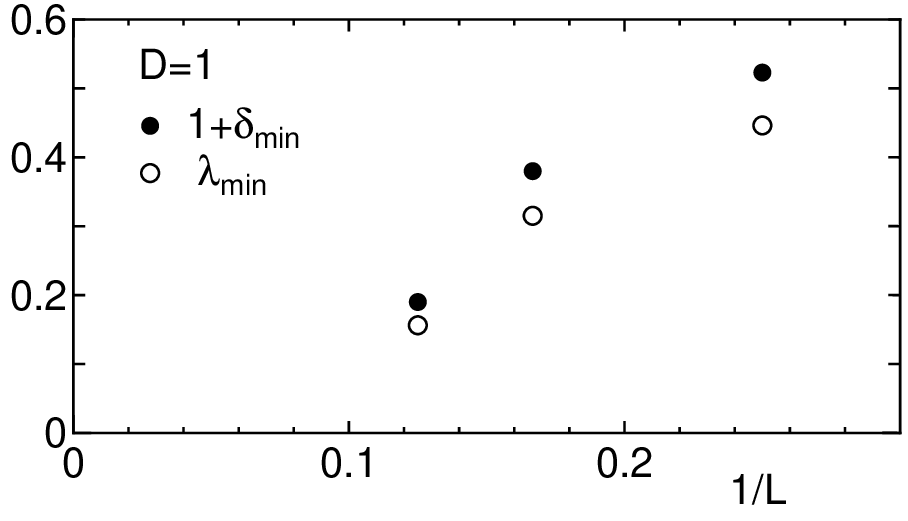}}

\caption{Minimum value of $1+\delta$ and $\lambda$ plotted against $1/L$ in the narrow N\'eel phase for $D=1$. }

\label{fig:dlmin}

\end{figure}

For large $\lambda$, there exists a wide N\'eel phase in the middle of the phase diagram, and two TLL phases are present on both sides of the N\'eel phase as discussed in \Sec\ref{sec:lambdainf}. In addition, another extremely narrow N\'eel region is found within the TLL phase. From the finite-size data, the TLL phases above and below this narrow N\'eel phase appear to be deformed into each other without passing through a phase boundary. However, the system size dependence of the end point $(\delta_{\rm min},\lambda_{\rm min})$ of this phase presented in Fig. \ref{fig:dlmin} suggests that $(\delta_{\rm min},\lambda_{\rm min})$ tends to $(\delta,\lambda)=(-1,0)$ for large enough system size. Hence, it is reasonable to conclude that these two TLL phases are separated  down to $(\delta,\lambda)=(-1,0)$. and the TLL phases above and below this narrow N\'eel phase correspond to the TLL0 and TLL1 phases described in  \Sec\ref{sec:deltam1} and \Sec\ref{sec:lambda0}, respectively. Furthermore, the TLL-LM1 phase boundary for $D=0$ obtained in ref.\citen{kh2024} plotted by the dotted line in Fig. \ref{fig:phasep}(c) near the point $(\delta,\lambda)=(-1,0)$ almost coincides with the narrow N\'eel phase that separates the TLL0 and TLL1 phases for $D=1$. Hence, the locations of the TLL0 and TLL1 phases are insensitive to the value of $D$. This indicates that the TLL0 and TLL1 phases originate from the TLL and LM1 phases of the isotropic model, respectively. 

\subsection{Case of easy axis anisotropy : $D=-1$}

\begin{figure}[h]

\centerline{\includegraphics[width=6.0cm]{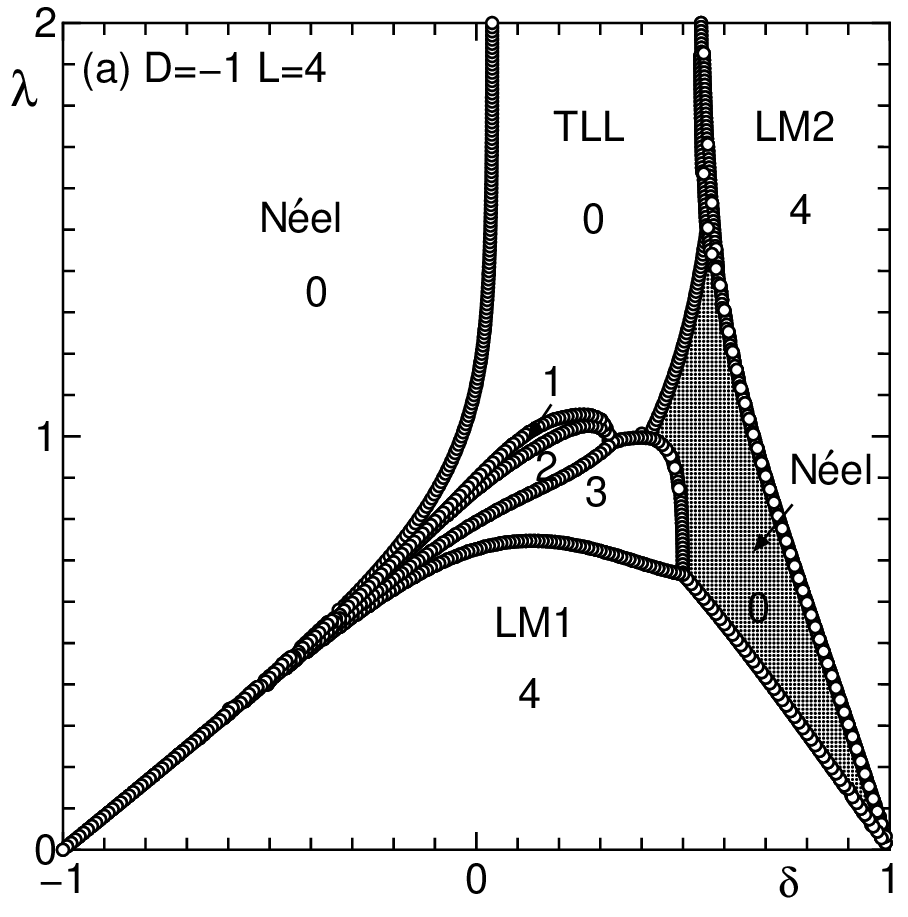}}

\centerline{\includegraphics[width=6.0cm]{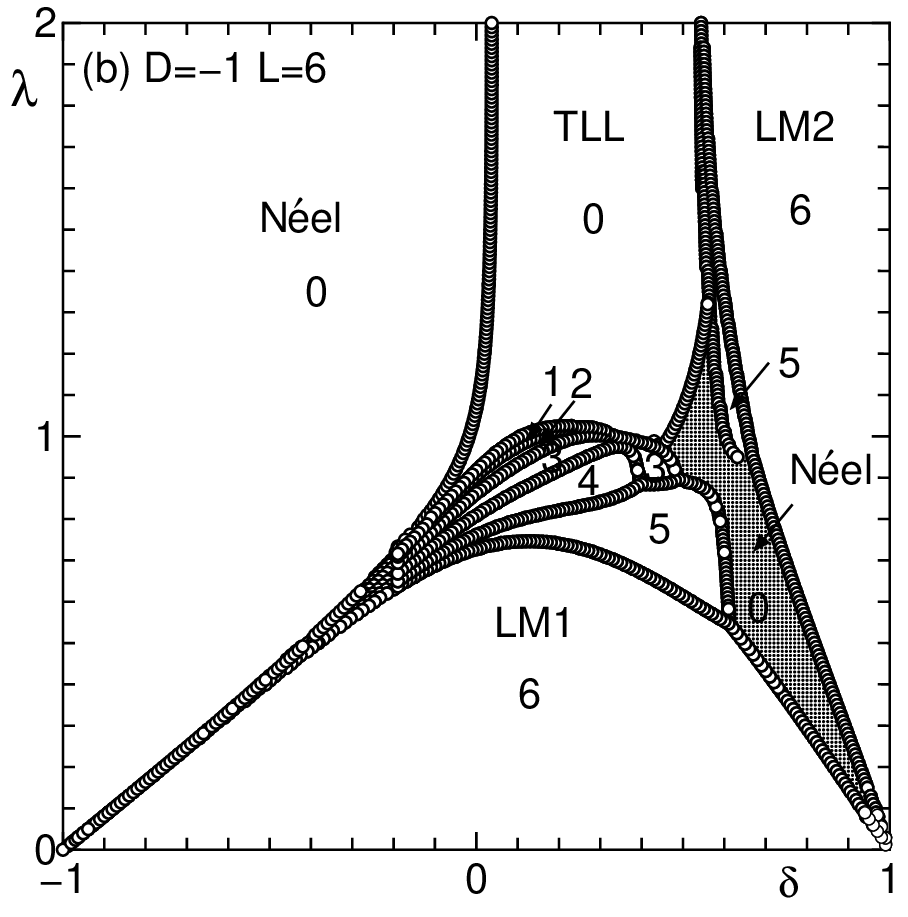}}

\centerline{\includegraphics[width=6.0cm]{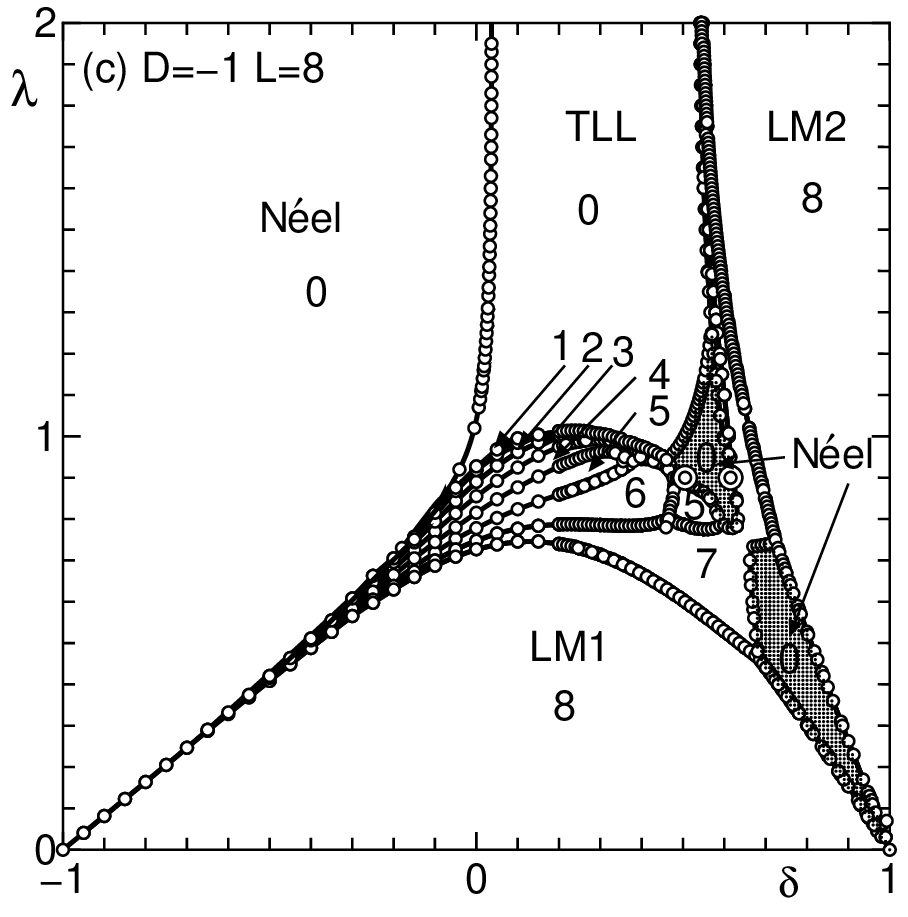}}

\caption{Ground-state phase diagrams based on the NED data with $D=-1$ for (a) $L=4$, (b) $L=6$, and (c) $L=8$. The number indicated for each phase is $\Mspz$ of the ground state. The double circles at $\lambda=0.9$ in (c) are the boundary of the nonmagnetic phase estimated from the DMRG calculation for $L=48$.}

\label{fig:phasem}

\end{figure}

We have carried out the NED calculation for $L=4$, 6, and 8 to obtain the ground-state phase diagram of Fig. \ref{fig:phasem} for $D=-1$. Again, around the center of the phase diagram, there are several ferrimagnetic phases that are expected to form a PF phase in the thermodynamic limit. The N\'eel phase in the shaded region of Fig. \ref{fig:phasem}(a)-(c) shrinks with the system size. Among them, for $L=8$, this N\'eel phase separates into two parts. In order to investigate the behavior for larger systems, we employ the DMRG method with open boundary conditions for $\lambda=0.9$ and 0.3. For $\lambda=0.9$, the width of the N\'eel phase is almost equal to that for $L=8$, while for $\lambda=0.3$, no phase with $\msp^z=0$ is observed. Hence, in the thermodynamic limit, we expect that the N\'eel phase with small $\lambda$ will be replaced by the PF phase with $0 <\msp <1$ as in the isotropic case.\cite{kh2024}.

\section{Summary and Discussion}

The ground-state phase diagram of mixed diamond chains with the Hamiltonian (\ref{eq:ham0}) is determined using the NED and the DMRG method in addition to the analytical approximations in various limiting cases. The phase diagram consists of a N\'eel ordered phase, nonmagnetic TLL phases, QF, and PF phases.

A region with anisotropy inversion is found where the Ising-like N\'eel phase is realized for easy plane anisotropy $D >0$ and the XY-like TLL phase is realized for easy axis anisotropy $D <0$ on the $S=1$ sites. The interpretation based on the effective Hamiltonian in the large $\lambda$ limit is given for this phenomenon. Similar inversion phenomena have been also found in one-dimensional frustrated $S = 1/2$ systems such as $S = 1/2$ frustrated 3-leg ladders\cite{okamoto2014},$S = 1/2$ distorted diamond chains\cite{oi2002,tokuno2005}, and $S=1/2$ trimerized spin chains with the next-nearest-neighbor interactions\cite{okamoto2002}. In these $S=1/2$ systems, the interplay of the frustration, the trimer nature of $S=1/2$ spins, and the anisotropic exchange interaction leads to the inversion phenomenon. In the present case, the pair of apical spins $\v{\tau}^{(1)}_l$ with magnitude 1/2 and $\v{\tau}^{(2)}_l$ with magnitude 1 would play the role of the $S=1/2$ trimer, since the $S=1$ spin can be regarded as a sum of two spins with magnitudes 1/2. It should be noted that the single-site anisotropy on the $S=1$ spin plays the role of an exchange anisotropy between these two spins with magnitudes $1/2$ forming $\v{\tau}^{(2)}_l$. In this context, the effect of exchange anisotropy on the present system would also be an interesting problem. This is left for future studies.

We found three different types of TLL phases for $D >0$. In the present work, they are distinguished by the ground-state phases of the isotropic model from which they emerge. However, it is desirable to characterize each of them by an appropriate order parameter or symmetry. This is also left for future studies.

In the present work, we focused on the cases of moderate strength of anisotropy $D=\pm 1$ to illustrate the typical picture of roles played by the single-site anisotropy $D$ on the ground state. However, the case of large $|D|$ would have a wider variety of features that are not observed in the moderate $D$ cases. As discussed above, the exchange anisotropy would also induce an even wider variety of phases. The further exploration of these effects is also left for future studies.

\acknowledgments
The numerical diagonalization program is based on the TITPACK ver.2 coded by H. Nishimori. The numerical computation in this work has been carried out using the facilities of the Supercomputer Center, Institute for Solid State Physics, University of Tokyo, and Yukawa Institute Computer Facility at Kyoto University.

\end{document}